\documentclass[10pt,a4j]{article}
\usepackage[dvips]{graphicx}
\usepackage{mathrsfs}
\usepackage{amsthm} 
\usepackage{amsmath}
\usepackage{amssymb}
\usepackage{amsfonts}
\usepackage{yhmath}
\usepackage{wrapfig}
\usepackage[dvips]{color}
\usepackage{cite}
\usepackage{array}
\usepackage{tabularx}
\usepackage[sectionbib]{chapterbib}
\usepackage{threeparttable}
\usepackage{chemarrow}
\usepackage{ascmac}
\usepackage{framed}
\definecolor{shadecolor}{gray}{0.90}

\begin{document}

\renewcommand{\figurename}{\small{Fig.}~}
\renewcommand{\labelitemi}{}
\renewcommand{\thefootnote}{$\dagger$\arabic{footnote}}
\renewcommand{\footnoterule}{%
  \vspace{2pt}
\flushleft\rule{6.154cm}{0.4pt}
  \vspace{4pt}
}
\pagestyle{plain}

\begin{flushright}
\textit{Coil Dimensions of Polystyrene Chain}
\end{flushright}
\vspace{0mm}

\begin{center}
\setlength{\baselineskip}{25pt}{\LARGE\textbf{Coil Dimensions as a Function of Concentration}}
\end{center}
\vspace{-5mm}
\begin{center}
\setlength{\baselineskip}{25pt}{\large\textbf{Coil Expansion of Polystyrene Chains}}
\end{center}
\vspace*{4mm}
\begin{center}
\large{Kazumi Suematsu} \vspace*{2mm}\\
\normalsize{\setlength{\baselineskip}{12pt} 
Institute of Mathematical Science\\
Ohkadai 2-31-9, Yokkaichi, Mie 512-1216, JAPAN\\
E-Mail: suematsu@m3.cty-net.ne.jp,  Tel/Fax: +81 (0) 593 26 8052}\\[5mm]
\end{center}@

\hrule
\vspace{0mm}
\begin{flushleft}
\textbf{\large Abstract}
\end{flushleft}
The preceding theory of excluded volume effects is applied to the Daud and coworkers$'$ observations. Based on various researchers' experimental data, it is suggested that the Daud and coworkers$'$ value of $\langle s^{2}\rangle^{1/2}$ at $\bar{\phi}=1$ may be revised from $82$ \text{\AA} to 93 \text{\AA}. Then agreement between the theory and the revised data is excellent, giving a support to the preceding result that the excluded volume effects should vanish at medium concentration.
\begin{flushleft}
\end{flushleft}
\hrule
\vspace{3mm}
\setlength{\baselineskip}{13pt}
\section{Introduction}
In this paper, the preceding theory\cite{Kazumi} of the excluded volume effects is reexamined. According to the theory, the expansion factor $\alpha$ is given by the equation:
\begin{equation}
\alpha^{5}-\alpha^{3}=N^{2}\frac{V_{2}^{\,2}}{V_{1}}\left(1/2-\chi\right)\left(\frac{\beta}{\pi}\right)^{3}\iiint\left(G_{hill}^{\,2}-G_{valley}^{\,2}\right)dxdydz\label{1-1}
\end{equation}
where $N$ is the number of segments, $V$ the volume (the subscripts 1 and 2 signifying solvent and segment, respectively), $\chi$ the enthalpy parameter defined by $\Delta H\propto\chi$, and $\beta=3/2\langle s^{2}\rangle_{0}$ ($\langle s^{2}\rangle_{0}$ being the mean square radius of gyration of an unperturbed chain).

In eq. (\ref{1-1}), $G$ is a function associated with segment concentration at the point $(x, y, z)$ (the subscripts $hill$ and $valley$ signifying concentrated and dilute regions, respectively) and has the form
\begin{equation}
G(x,y,z)=\sum_{\{a,b,c\}}\exp\{-\beta[(x-a)^{2}+(y-b)^{2}+(z-c)^{2}]\}\label{1-2}
\end{equation}

\begin{wrapfigure}[14]{r}{7cm}
\vspace*{-6mm}
\begin{center}
\includegraphics[width=6.5cm]{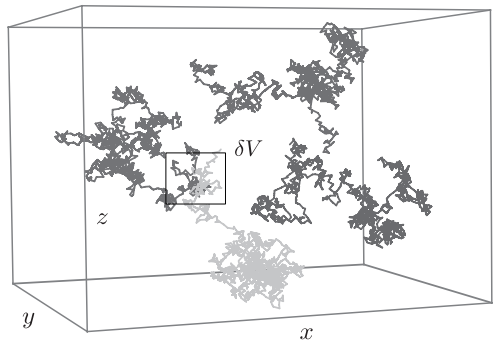}
\end{center}
\caption{A snapshot of a polymer solution.}\label{snapshot}
\end{wrapfigure}

\noindent where $\{a,b,c\}$ is a set of coordinates of the center of gravity of polymer molecules in the solution, so that the summation represents the accumulation of segments emanating from different polymer chains and hence $G$ reflects the segment concentration at the coordinate $(x,y,z)$.

Eq. (\ref{1-1}) suggests a new aspect of polymer solutions: The expansion factor $\alpha$ is a function of the difference $\varDelta=G_{hill}^{\,2}-G_{valley}^{\,2}$, so $\alpha$ is a strong function of concentration fluctuation.

A polymer solution is in itself an inhomogeneous solution of segments (see Fig. \ref{snapshot}). This is because monomers are joined by chemical bonds; for this reason the segment concentration is the highest at the center of gravity, but decreases monotonically with increasing distance $s$ from the center.

Such large inhomogeneity, however, can be reduced by raising polymer concentration, since the interpenetration of segments develops progressively with increasing concentration. According to eq. (\ref{1-1}), as the inhomogeneity $\varDelta$ decreases, the excluded volume effects must also decrease. Thus the problem of evaluating the coil dimensions reduces to the problem of evaluating the magnitude of the inhomogeneity as a function of polymer concentration.

\section{Application to Experiments}
Small angle neutron scattering (SANS) measurements give valuable information on the mean square radius of gyration, $\langle s^{2}\rangle$, of polymer chains over wide concentration range\cite{Daud, Graessley, Westermann, Cotton, Wignall-1, Wignall-2, Tangari}. Thus the SANS experiment provides the best tool to examine the theory of the excluded volume effects. In this paper we take up the earliest experiment by Daud and coworkers\cite{Daud} (polystyrene solution in CS$_{2}$; MW=114000; M$_{w}$/M$_{n}\cong\,$1.1) to compare with the preceding theory.

We introduce the lattice model to calculate the integral term:
\begin{equation}
J=\iiint\left(G_{hill}^{\,2}-G_{valley}^{\,2}\right)dxdydz\label{1-3}
\end{equation}
According to the preceding work, polymer molecules are arranged on the sites of the simple cubic lattice having the unit length $p\times p\times p$. Then $G$ has the form:
\begin{equation}
G(x,y,z)=\sum_{i=1}^{\infty}\sum_{j=1}^{\infty}\sum_{k=1}^{\infty}\exp\{-\beta[(x-ip)^{2}+(y-jp)^{2}+(z-kp)^{2}]\}\label{1-4}
\end{equation}
so that the coordinate $(ip, jp, kp)$ represents the location of the center of gravity of a polymer molecule. Let us define the hill area as a region enclosed by the interval $[-p/4, p/4]$ for each axis and the valley area as that enclosed by $[p/4, 3p/4]$. Then eq. (\ref{1-3}) may be recast in the form:
\begin{equation}
J=\iiint_{-p/4}^{p/4} G^{\,2}dxdydz-\iiint_{p/4}^{3p/4} G^{\,2}dxdydz\label{1-5}
\end{equation}
The average segment density can be calculated by the equation:
\begin{equation}
\bar{\phi}=\frac{V_{2}N}{p^{3}}\label{1-6}
\end{equation} 
With the help of eqs. (\ref{1-4}), (\ref{1-5}) and (\ref{1-6}), we can solve eq. (\ref{1-1}) as a function of $\bar{\phi}$.

The basic physicochemical parameters of polystyrene are listed in Table \ref{Daud}. To date the enthalpy parameter $\chi$ for carbon disulfide (CS$_{2}$) remains unknown. There are some evidences\cite{Yamamoto, Matsushita, Fukuda, Cotton}, however, that CS$_{2}$ will have slightly smaller value of $\chi$ than that for benzene\footnote{\,\, $\chi=0.44\sim 0.4$ for $\bar{\phi}=0\sim 0.4$ at 25$^{\circ}$C\cite{Barton}.}. Taking these data into consideration, we have chosen $\chi=0.4$.

A central problem throughout the present work is the coil size in the bulk state ($\bar{\phi}=1$). It was pointed out\cite{Westermann} recently that the observed value by Daud and coworkers\cite{Daud} is not consistent with the observations by other workers. To inspect this we have collected experimental data of various researchers\cite{Flory, Berry, Yamamoto, Matsushita, Fukuda, Cotton, Wignall-1, Wignall-2, Tangari, Fetters} and compared those with the Daud and coworkers$'$, the results being summarized in Table A (see Appendix). As one can see, the Daud and coworkers$'$ evaluation ($82$ \text{\AA}) is appreciably smaller than the other researchers$'$ (the average of which is $93.4$ \text{\AA}). On the basis of this information, we replace the original value $82$ \text{\AA} with $93$ \text{\AA} (see Table \ref{observed}) and compare with eq. (\ref{1-1}).

\begin{table}[!htb]
\vspace{0mm}
\caption{Basic parameters of polystyrene solution\label{Daud}}
\begin{center}
\vspace*{-1.5mm}
\begin{tabular}{l l c r}\hline\\[-1.5mm]
& \hspace{10mm}parameters & notations & values \,\,\,\,\\[2mm]
\hline\\[-1.5mm]
polystyrene (PSt) & volume of a solvent (CS$_{2}$) & $V_{1}$ & \hspace{5mm}100 \text{\AA}$^{3}$\\[1.5mm]
& volume of a segment (C$_{8}$H$_{8}$) & $V_{2}$ & \hspace{5mm}165 \text{\AA}$^{3}$\\[1.5mm]
& degree of polymerization & $N$ & \hspace{5mm}1096 \,\,\,\,\,\,\,\\[1.5mm]
& Flory characteristic ratio & C$_{F}$ & \hspace{5mm}10 \,\,\,\,\,\,\,\\[1.5mm]
& mean bond length & $\bar{\ell}$ & \hspace{5mm}1.55 \text{\AA}\,\,\,\\[1.5mm]
& enthalpy parameter (25$^{\,\circ}$C) & $\chi$ & \hspace{5mm}0.4 \,\,\,\,\,\,\,\\[2mm]
\hline\\[-6mm]
\end{tabular}\\[6mm]
\end{center}
\end{table}

\begin{figure}[ht]
\begin{center}
\begin{minipage}[t]{0.53\textwidth}
\begin{center}
\includegraphics[width=9cm]{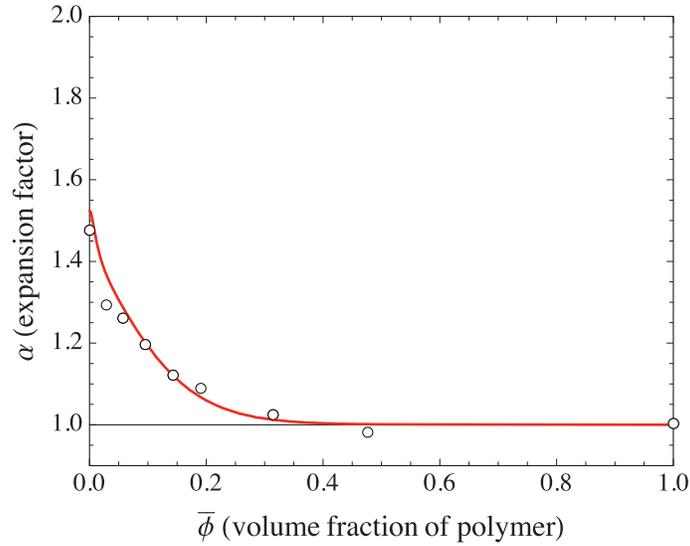}
\end{center}
\vspace{-2mm}
\caption{Expansion factor as against the average volume fraction $\bar{\phi}$ of polymer. Solid line (\textcolor{red}{$-$}): theoretical line by eq. (\ref{1-1}). Open circles ($\circ$): observed points by Daud and coworkers\cite{Daud} along with the revised point at $\bar{\phi}=1$ (the observed value of 82 \text{\AA} at $\bar{\phi}=1$ was replaced by the revised one 93 \text{\AA}, so that the original point should have the location of $\alpha=0.89$ at $\bar{\phi}=1$).}\label{polystyrene}
\end{minipage}
\end{center}
\end{figure}

The simulation result is illustrated in Fig. \ref{polystyrene}, together with the Daud and coworkers$'$ observed points ($\circ$) and the revised point 93 \text{\AA} at $\bar{\phi}=1$. Agreement between the theory (solid line) and the experiments is remarkably good, giving a support to the preceding result\cite{Kazumi} that the excluded volume effects should vanish at medium concentration. The present result is in contrast to the classic prediction that the excluded volume effects should decrease smoothly over the entire concentration range from the dilution limit to the melt, but in accord with the experimental observations by Cheng, Graessley, and Melnichenko\cite{Graessley}, and those by Westerman, Willner, Richter and Fetters\cite{Westermann}.\\

\begin{center}
  \begin{threeparttable}
    \caption{Observed radii of gyration of polystyrene chains\cite{Daud}}\label{observed}
  \begin{tabular}{l l r l}
  \hline\\[-1.5mm]
  \hspace{2mm}molecular weight (M$_{\text{W}}$) & \hspace{12mm}$\bar{\phi}$ & \hspace{5mm}$\langle s^{2}\rangle^{1/2}$  (\text{\AA})\\[2mm]
\hline\\[-2.5mm]
\hspace{10mm}114000 & \hspace{10mm}0.000 & \hspace{5mm}137\,\,\,\,\,\,\,\,\,\\[1mm]
& \hspace{10mm}0.028 & \hspace{5mm}120\,\,\,\,\,\,\,\,\,\\[1mm]
& \hspace{10mm}0.057 & \hspace{5mm}117\,\,\,\,\,\,\,\,\,\\[1mm]
& \hspace{10mm}0.095 & \hspace{5mm}111\,\,\,\,\,\,\,\,\,\\[1mm]
& \hspace{10mm}0.143 & \hspace{5mm}104\,\,\,\,\,\,\,\,\,\\[1mm]
& \hspace{10mm}0.190 & \hspace{5mm}101\,\,\,\,\,\,\,\,\,\\[1mm]
& \hspace{10mm}0.314 & \hspace{5mm}95\,\,\,\,\,\,\,\,\,\\[1mm]
& \hspace{10mm}0.476 & \hspace{5mm}91\,\,\,\,\,\,\,\,\,\\[1mm]
& \hspace{10mm}1.000 & \hspace{5mm}93\tnote{\,a}\,\,\,\,\,\,\,\,\,\\[1mm]
\hline\\[-6mm]
   \end{tabular}
    \vspace*{2mm}
   \begin{tablenotes}
     \item a. Original value is 82 \text{\AA}.
   \end{tablenotes}
  \end{threeparttable}
\end{center}

\vspace{4mm}
\vspace*{0mm}

\section{Discussion}
A fundamental question is why a polymer coil expands. The answer seems quite simple: it simply comes from the difference of the Gibbs potential between the inside and the outside of a polymer coil, as the classic thermodynamics states clearly that a chemical equilibrium is realized at $\Delta G=0$ under constant $T$ and $P$. The potential in question is caused by the concentration gradient between the inside and the outside of a molecule, namely by the inhomogeneity just mentioned above. Now it becomes clear that the wild inhomogeneity intrinsic to polymer solutions is the very source of the coil expansion. Theorists have so far neglected this large inhomogeneity of polymer solutions, having resorted to the mean-field ansatz. 

It is important to stress that the present result is by no means conclusive. The agreement seen in Fig. \ref{polystyrene} is based on the revised data, namely 82 \text{\AA} $\rightarrow$ 93 \text{\AA} for $\bar{\phi}=1$. To settle the long-standing problem in polymer physics, the concentration dependence of excluded volume effects, much more experimental investigations are necessary.\\

\section{Appendix}
\begin{center}
  \begin{threeparttable}
    \centering{Table A: Estimation of unperturbed radii of gyration for polystyrene}\\[2mm]
  \begin{tabular}{c r c l l}
    \hline\\[-1.5mm]
No. & \hspace{3mm}M$_{\text{W}}$\hspace{2.5mm} & \hspace{3mm}$\langle s^{2}\rangle_{0}^{1/2}$ (\AA) & Instruments \& Empirical Equations & \hspace{2mm}Literature \\[2mm]
\hline\\[-1.5mm]
1 & 114000 & \hspace{3mm}82 \text{\AA} & \hspace{5mm}SANS\tnote{\,a}\,\, (bulk) & \hspace{2mm}Daud et al.\cite{Daud}\\[2mm]
\hline\\[-1.5mm]
2 & \hspace{3mm}114000 & \hspace{3mm}93.5 \text{\AA} & \hspace{5mm}RIS\tnote{\,b} & \hspace{2mm}Flory\cite{Flory}\\[1.5mm]
& & & \hspace{5mm}$\langle s^{2}\rangle_{0}^{1/2}=\left(\frac{1}{6}\,C_{F}\,n\,\ell^{2}\right)^{1/2}$\hspace{3mm}$(C_{F}=10)$ & \\[2mm]
\hline\\[-1.5mm]
3 & 114000 & \hspace{3mm}93.2 \text{\AA} & \hspace{5mm}LS\tnote{\,c}\,\, (cyclohexane \& decalin) & \hspace{2mm}Berry\cite{Berry}\\[1mm]
& & & \hspace{5mm}$\langle s^{2}\rangle_{0}^{1/2}=0.276\,M^{1/2}$ (\text{\AA}) & \\[2mm]
\hline\\[-1.5mm]
4 & 114000 & \hspace{3mm}97.8 \text{\AA} & \hspace{5mm}LS (cyclohexane) & \hspace{2mm}Yamamoto et al.\cite{Yamamoto, Matsushita}\\[1mm]
& & & \hspace{5mm}$\langle s^{2}\rangle_{0}^{1/2}=0.290\,M^{1/2}$ (\text{\AA}) & \\[2mm]
\hline\\[-1.5mm]
5 & 114000 & \hspace{3mm}94.5 \text{\AA} & \hspace{5mm}LS (\textit{trans}-decalin) & \hspace{2mm}Fukuda et al.\cite{Fukuda}\\[1mm]
& & & \hspace{5mm}$\langle s^{2}\rangle_{0}^{1/2}=0.28\,M^{1/2}$ (\text{\AA}) & \\[2mm]
\hline\\[-1.5mm]
6 & \hspace{3mm}114000 & \hspace{3mm}92.9 \text{\AA} & \hspace{5mm}SANS (bulk) & \hspace{2mm}Cotton et al.\cite{Cotton}\\[1.5mm]
& & & \hspace{5mm}$\langle s^{2}\rangle_{0}^{1/2}=0.275\,M^{1/2}$ (\text{\AA}) &  \\[2mm]
\hline\\[-1.5mm]
7 & 97200 & \hspace{3mm}84.0 \text{\AA}\tnote{d} & \hspace{5mm}SANS (bulk) & \hspace{2mm}Wignal et al.\cite{Wignall-1,Wignall-2} \\[1.5mm]
& 114000 & \hspace{3mm}91.2 \text{\AA} & \hspace{5mm}$\langle s^{2}\rangle_{0}^{1/2}=0.27\,M^{1/2} $ (\text{\AA}) &  \\[2mm]
\hline\\[-1.5mm]
8 & 114000 & \hspace{3mm}90.1 \text{\AA} & \hspace{5mm}SANS (bulk) & \hspace{2mm}Tangari et al.\cite{Tangari}\\[1.5mm]
& & & \hspace{5mm}$\langle s^{2}\rangle_{0}^{1/2}=0.267\,M^{1/2}$ (\text{\AA}) &  \\[2mm]\hline\\[-1.5mm]
9 & 114000 & \hspace{3mm}94.2 \text{\AA} & \hspace{5mm}LS (cyclohexane) & \hspace{2mm}Fetters et al.\cite{Fetters}\\[1mm]
& &  & \hspace{5mm}$\langle s^{2}\rangle_{0}^{1/2}=0.279\,M^{1/2}$ (\text{\AA}) & \\[2mm]
\hline\\[-6mm]
   \end{tabular}
    \vspace*{2mm}
   \begin{tablenotes}
     \item a. Small angle neutron scattering
     \item b. Rotational isomeric state model
     \item c. Light scattering
     \item d. The average value of the observed points: $80, 84, 86, \text{and}\, 87$ \AA.
   \end{tablenotes}
  \end{threeparttable}
\end{center}

\newpage

\end{document}